\documentclass{ws-procs9x6}
\usepackage{graphicx}
\newcommand{\be}{\begin{eqnarray}}
\newcommand{\ee}{\end{eqnarray}}
\begin{document}

\title{STRONGLY COUPLED QUARK-GLUON PLASMA: \\ THE STATUS REPORT}

\author{E.V.Shuryak$^*$ }

\address{ Department of Physics and Astronomy, University at Stony Brook,\\
Stony Brook NY 11794 USA\\
$^*$E-mail: shuryak@tonic.physics.sunysb.edu}

\begin{abstract}
%Understanding of sQGP is progressing very repidly, and yet
%we are far from really having it. 
 RHIC data have shown robust collective flows,
strong jet and charm quenching, and charm flow.
 Recently ``conical flows'' from damped jets were seen.
Non-Abelian classical strongly coupled plasmas were
introduced and studied via molecular dynamics, 
with first results
for its transport (diffusion and viscosity)  reported.
 Quantum-mechanical studies reveal the survival for $T>T_c$
of the lowest binary states, including colored ones, and
 also of some manybody ones such as  baryons.
``Polymeric
chains'' $\bar q.g.g... q$ are also bound in some range of $T$,
perhaps the  progenitors of the QCD strings.
 AdS/CFT applications advanced to a completely new level of detail:
they now include studies of thermal  heavy quark
motion, jet quenching and even conical flow.
Confinement is however still beyond simply strong coupling:
its de-facto inclusion the so called
AdS/QCD approach is so far added as a model, while true understanding
may probably only come from further insights
into the monopole dynamics.
\end{abstract}


\bodymatter

\section{Why strongly coupled?}

 A realization~\cite{Shu_liquid,SZ_rethinking,SZ_CFT} that
QGP at RHIC  is not a weakly coupled 
gas but rather a strongly coupled liquid
 has lead to a  paradigm
 shift in the field. It was extensively debated  at
the ``discovery''  BNL workshop in 2004~\cite{discovery_workshop} (at which the
abbreviation sQGP was established) and multiple other meetings since.

In the intervening three years we
had to learn a lot, some new some 
 from other branches of physics  which
happened to have some experience with strongly coupled systems.
Those range from
quantum gases to classical plasmas to string theory.
In short, there seem to be not one
but  actually two difficult
issues we are facing. One is to understand why QGP at $T\sim 2T_c$
is strongly coupled, and what exactly it means. The second large
problem is to understand what  happens
 at the deconfinement, at $|T-T_c| \ll T_c$,
which may be a key to the famous confinement problem. 

As usual, progress proceeds from
 catching/formulating the main concepts and qualitative pictures, to
mastering technical tools, to final quantitative
predictions: and now we are somewhere in the middle
of this process. The work is going on at many fronts.
At classical level, 
 first studies of the  transport properties
of strongly coupled non-Abelian plasmas have been made.
 Quantum-mechanical studies
of the bound states above $T_c$ have revealed a lot of unusual
states, including ``polymeric chains''.  At the quantum field/string 
theory front, a surprisingly detailed uses of
 AdS/CFT correspondence has been made.
 And yet, to be honest, 
 deep  understanding   is still missing: e.g.
we don't know what the CFT plasma is made of.

The list of arguments explaining why we think  QGP
is strongly coupled at $T$ above $T_c$
is long and constantly growing. Let me start with its
short version, as I see them today.\\
\noindent
1.Collective phenomena observed at RHIC lead
 hydro practitioners to a conclusion that
 QGP as a ``near perfect liquid'', with unusually
small {\em viscosity-to-entropy ratio}
$\eta/s=.1-.2<<1$ \cite{Teaney:2003kp}
in striking contrast to pQCD predictions.
Not only light jets, but also charmed ones are strongly quenched.
Charm diffusion constant $D_c$ deduced from its flow is 
 an order of magnitude lower than pQCD estimates
\cite{MT}. 
\\
2. Combining lattice data on quasiparticle masses and
interparticle
potentials, one  finds a lot of quasiparticle bound states
\cite{SZ_bound}. 
The same approach explains why $\eta_c,J/\psi$ remains bound
till near $3T_c$, as was directly observed on the lattice
\cite{charmonium} and perhaps experimentally at RHIC.
The resulting resonances
enhance transport cross sections \cite{SZ_rethinking,Rapp_vanHees}
and may lead to a liquid-like behavior. Similar thing 
does happen for ultracold trapped atoms, due to
Feshbach-type resonances at which
the scattering length $a\rightarrow \infty$.
\\
3.The interaction parameter  
$\Gamma\sim <potential\, energy>/<kinetic\, energy>$
in sQGP is obviously not small. Classical
  e/m plasmas  at the comparable
coupling $\Gamma\sim 1-10$ are  good liquids too.
%and results to be described below show it to be so for non-Abelian
%plasmas as well.
\\
4. Exact correspondence between a conformal
(CFT) $\cal N$=4 supersymmetric Yang-Mills theory
at strong coupling and string theory in Anti-de-Sitter
space (AdS) in classical SUGRA regime was conjectured by Maldacena~\cite{Maldacena:1997re}.
 The results obtained this way on the 
 $g^2N_c\rightarrow \infty$ regime of the CFT plasma 
 are all  close to what we
 know about sQGP. Indeed, it has a very similar thermodynamics and
 is a good liquid with record low viscosity as well. Recent works
(see below)
added to the list  parametrically 
large jet quenching  and  small diffusion constant
for heavy quarks.
\\
5.\footnote{This part is presented at this conference for the first time.}
The \cal{N}=2 SUSY YM (``Seiberg-Witten'' theory) 
is a working example of confinement due to 
condensed monopoles\cite{SW}.
If it is also true for QCD,  at $T\rightarrow T_c$
magnetic monopoles must become light
 and weakly interacting at large distances due to U(1) beta function.
Then the Dirac condition forces
 electric coupling $g$  be large (in IR).

\section{Collective Flows in Heavy Ion Collisions}
 This meeting is mostly theoretical in nature, and thus
I would not go into details of heavy ion phenomenology.
Collective flows, related with explosive behavior of hot
matter, were
 observed at SPS and RHIC and are quite accurately reproduced
by the ideal hydrodynamics.
 The flow affect different secondaries differently, yet their
spectra are in quantitative agreement with the data
for all of them, from $\pi$ to $\Omega^-$. 
At non-zero impact parameter the
original excited system is deformed in the transverse plane,
creating the so called elliptic flow.
It is described by the parameter $v_2(s,pt,M_i,y,b,A)\\ =<cos(2\phi)>$,
where $\phi$ is the azimuthal angle and the others stand for
the collision energy, transverse momentum, particle mass, rapidity, centrality
and system size. Hydrodynamics explains well all of those dependences,
for about 99\% of the particles\footnote{At
large $p_t>2 GeV$  a different regime starts, related with jets.}.

\begin{figure}
 \includegraphics[width=4cm]{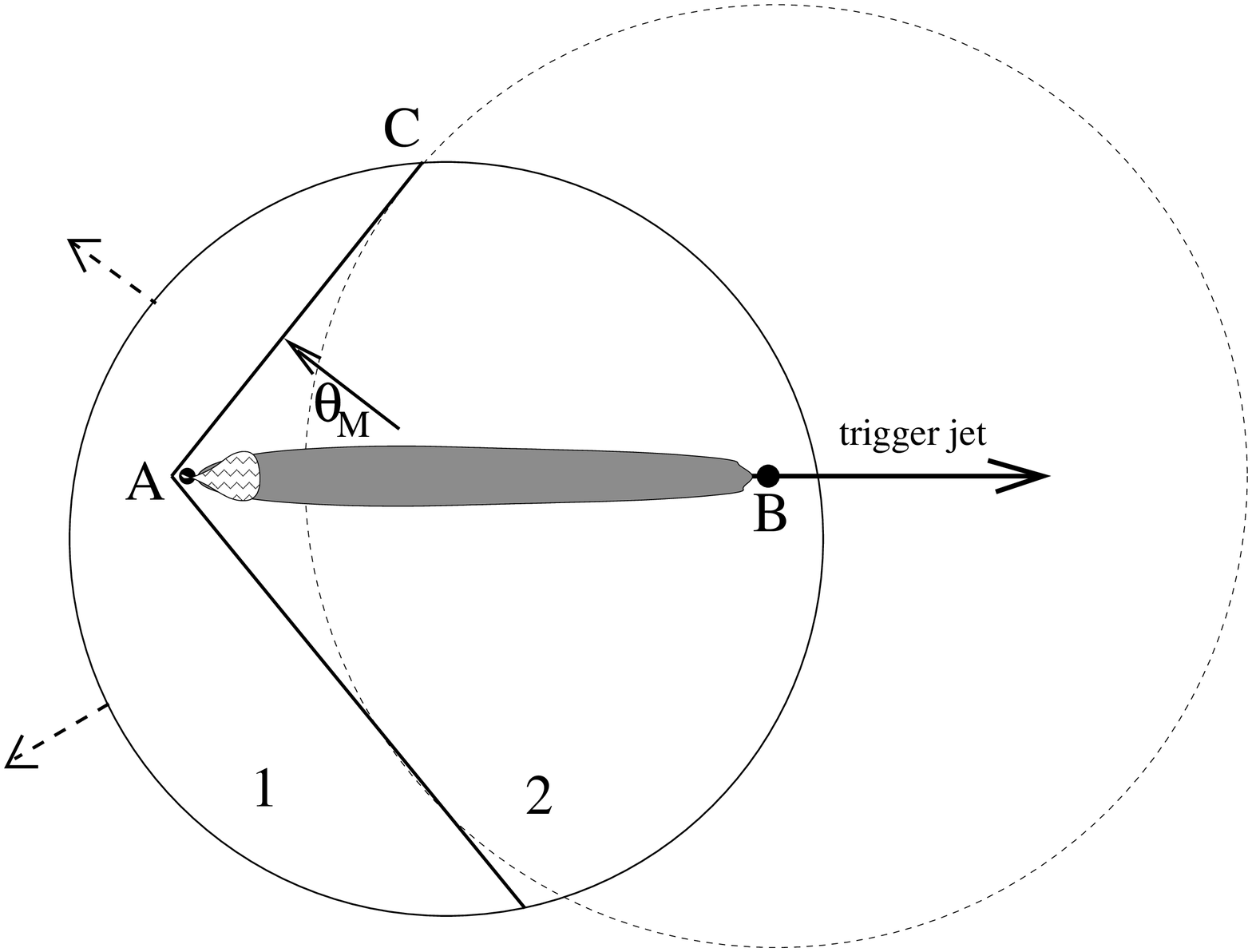}
 \includegraphics[width=6cm]{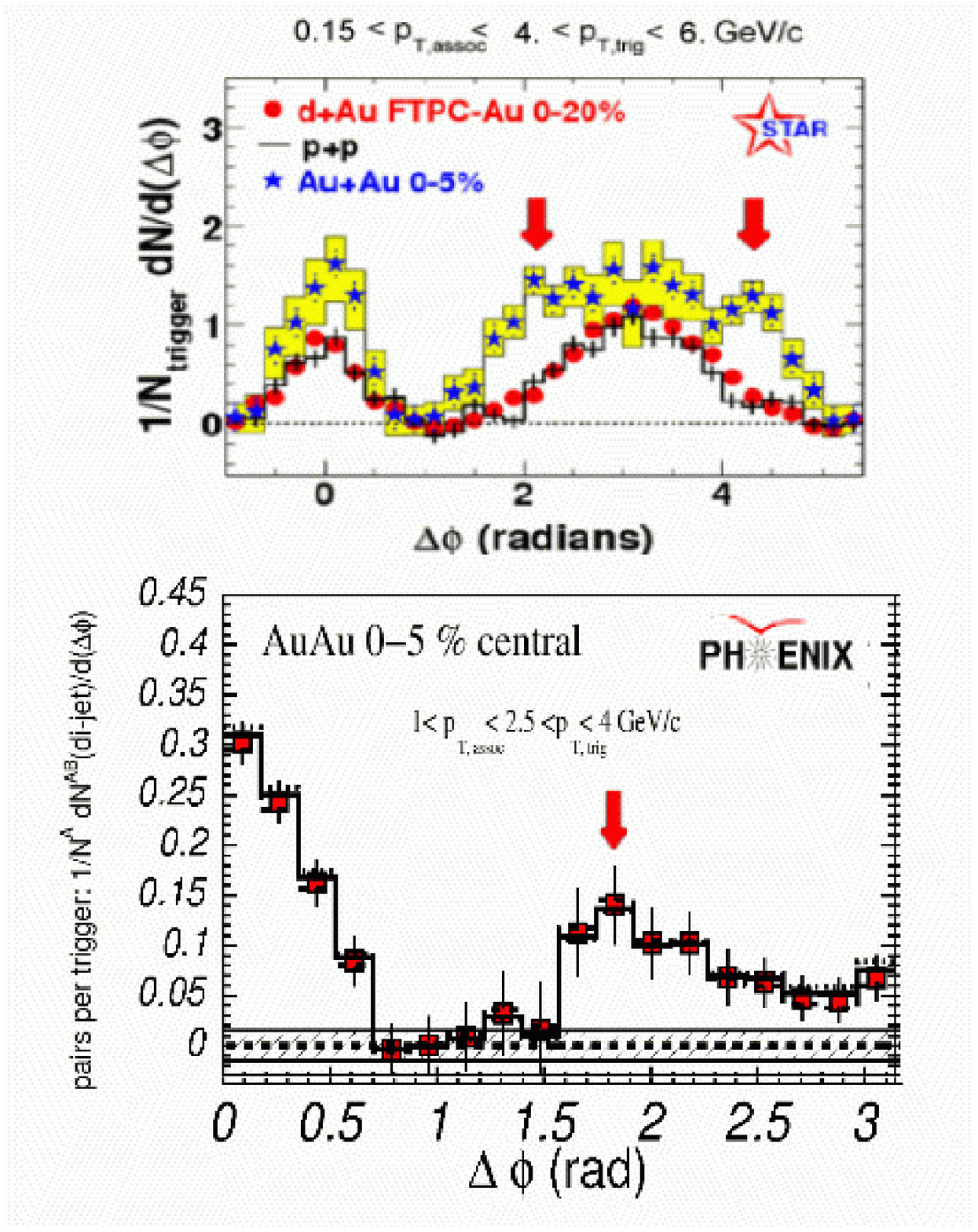}
 \caption{\label{fig_shocks}
(a) A schematic picture of flow created by a jet going through
the fireball. The trigger jet is going to the right
from the origination point  B. The
  companion quenched jet is moving to the left, heating the matter
(in shadowed  area) and producing a shock cone
%and thus creating a cylinder of additional matter (shaded area).
% The head of the jet is a ``non-hydrodynamical core'' of the QCD gluonic shower,
%formed by the original hard parton (black dot A).
with a flow normal to it, at
the Mach angle $cos\theta_M=v/c_s$, where $v,c_s$ are jet and sound velocities.
(b)The background subtracted correlation functions from STAR and PHENIX
experiments, a distribution in azimuthal
angle $\Delta\phi$ between the trigger jet and associated particle.
Unlike in pp and $dAu$ collisions where the decay of the companion jet
create a peak at  $\Delta\phi=\pi$ (STAR plot), central $AuAu$
collisions show a minimum at that angle and a maximum corresponding
to the Mach angle (downward arrows).
}
 \end{figure}

 New hydrodynamical phenomenon suggested recently \cite{CST},
is the so called {\em conical} flow which is induced by
jets quenched in sQGP. Although the QCD Lagrangian tells us that 
 charges are coupled to gluons and thus it is gluons
which are to be radiated, at strong coupling those are
rapidly quenched.  Effectively 
the jet energy is dumped into the medium and then
it transformed into  coherent radiation of
sound waves, which unlike gluons
 are  much less absorbed and can survive till freezout to be detected.
As shown in Fig.\ref{fig_shocks}, this seem to be what indeed is observed.

%  Fig.\ref{fig_shocks} explains a view of the process,
%in a plane transverse to the beam.  Two oppositely
%moving jets originate from   the hard collision point B.
% Due to strong quenching, the survival of the trigger
%jet biases it to be produced close to the surface and to
% move outward. This  forces its companion to 
%move inward through matter and to be maximally quenched.
%The energy deposition starts at point B, thus a spherical sound wave
%appears (the dashed circle in Fig.\ref{fig_shocks} ). Further 
% energy deposition is along the jet line, and is propagating with a speed of
%light, 
%till the leading parton is found at point A
%at the moment of the snapshot.

%The main prediction is that the shape of the jet passing through sQGP
%grastically changes: most of
%associated secondaries 
%  fly preferentially to a very large angle with jet direction,
% $\approx 70$ degrees
% consistent with the Mach angle for (
%a time-averaged) speed of sound.
  
Antinori and myself~\cite{Antinori:2005tu} suggested to test it further
by  b-quark jets,
which can be tagged experimentally even if not ultrarelativistic:
  the Mach
cone should then shrink, till it goes to zero at the critical velocity
$v=c_s=1/\sqrt{3}$. Gluon radiation behaves oppositely, expanding
 with $decreasing$
$v$, and never shrinks to zero.   
Casalderrey and myself\cite{CS_variable} have shown, using 
conservation of adiabatic invariants, that fireball expansion 
should greatly enhance the sonic boom\footnote{ The reason is similar to
enhancement
of a sea wave such tsunami as it goes onshore.}.

\section{Classical strongly coupled non-Abelian plasmas}

In the  electromagnetic plasmas
the term ``strongly coupled''  is expressed via
 dimensionless parameter $\Gamma= (Ze)^2/(a_{WS}T)$
characterizing the strength of the interparticle interaction.
 $Ze,a_{WS},T$ are respectively the ion charge, the Wigner-Seitz radius
$a_{WT}=(3/4\pi n)^{1/3}$  and the temperature. $\Gamma$ is convenient 
to use because it only involves the {\it input} parameters, such as the 
temperature and density. 
Extensive studies 
using both MD and analytical methods,
have revealed the following regimes:
{\bf i.} a gas regime for $\Gamma<1$; {\bf ii.} a
liquid regime for $\Gamma\approx  10$;  {\bf iii.} a glass regime
for $\Gamma\approx 100$; {\bf iv.} a solid regime for $\Gamma > 300$. 

 Gelman, Zahed and myself~\cite{GSZ}
 proposed a  model for the description of strongly interacting
quarks and gluon quasiparticles  as a classical
and nonrelativistic Non-Abelian Coulomb gas. The sign and strength
of the inter-particle interactions are fixed by the scalar product
of their classical {\it color vectors} subject to Wong's equations.
The EoM for the phase space coordinates follow from the usual 
Poisson brackets:
\be
\{  x_{\alpha\,i}^m, p_{\beta\,j}^n \}=\delta^{mn} 
\delta_{\alpha\beta}\delta_{ij}  \,\,\,\, \{ Q_{\alpha\,i}^a, Q_{\beta\,j}^b\}= f^{abc}\,Q_{\alpha\,i}^c
\ee
For the color coordinates they are 
classical analogue of the SU(N$_c$) color commutators,
with $ f^{abc}$ the  structure constants of the color group.
The classical color vectors are all adjoint vectors with
$a=1...(N_c^2-1)$. For  the non-Abelian group SU(2)
those are 3d vectors on a unit sphere, for SU(3) there are
8 dimensions minus 2 Casimirs=6 d.o.f.\footnote{
Although color EoM do not look like the usual
canonical relations between coordinates and momenta, they actually
are pairs of conjugated variables, as can be shown via
 the so called Darboux parameterization.}.

%%%%%

\begin{figure}[ht]
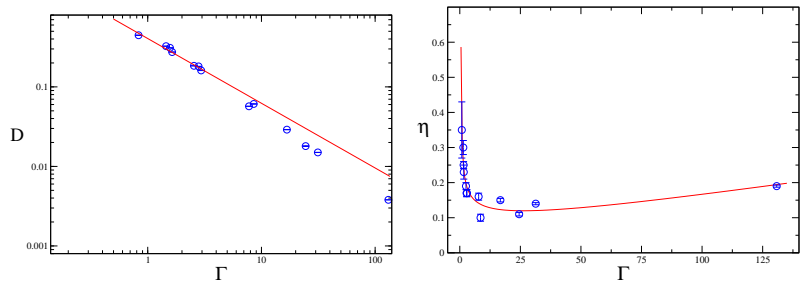

\begin{center}
\epsfig{figure=diffusion.eps,width=5.1cm,angle=0}\hspace{.2cm}
\epsfig{figure=viscosity.eps,width=5.1cm,angle=0}
\caption{The diffusion constant (a) and shear viscosity (b)
 of a one species cQGP as a function of
the dimensionless coupling $\Gamma$.  Blue points are the MD simulations; the
red curve is the fit.}
\label{fig_diffusion}
\end{center}
\end{figure}

The model was studied using Molecular Dynamics (MD),
which means solving numerically EoM for $n\sim 100$ particles.
It also displays a number of phases as the Coulomb coupling is
increased ranging from a gas, to a liquid, to a crystal with
anti-ferromagnetic-like color ordering. There is no place for details
here: in Fig.\ref{fig_diffusion} one can see the result for
 diffusion and viscosity vs coupling:
note how different and nontrivial they are. 
When extrapolated to the sQGP
suggest that the phase is liquid-like, with a diffusion constant 
$D\approx 0.1/T$ and a bulk viscosity to entropy density ratio 
$\eta/s\approx 1/3$.
The second paper of the same group\cite{GSZ} discussed the energy
and
the screening at $\Gamma>1$, finding large deviations from the Debye
theory.

\section{Quantum mechanics of the quasiparticles}

In the deconfined phase, at $T>T_c$, the basic objects are
dressed quarks and gluons. Even perturbatively 
they get masses $M_{eff}\sim g T$ \cite{Shu_QGP}
and dispersion curve close to that of massive particle.
However it follows from lattice 
measurements~\cite{masses} 
(admittedly, with still poor accuracy)  that (i) masses are very
large,
about $M_{eff}\approx (3-4) T$; and (ii) quarks and gluons have 
very close masses, in contrast to pQCD prediction. Thus, in first
approximation, quasiparticles are rather non-relativistic.

As emphasized by Zahed and myself~\cite{SZ_bound}, a gas of such heavy
quasiparticles cannot generate the pressure observed in other lattice
works. The resolution of a puzzle may be found if there are multiple
bound states of the quasiparticles, which also contribute to pressure.

The existence of bound states also follows from the interaction
deduced for static quarks. For marginal states, with near-zero
binding,
those should be applicable. The most obvious state to think of is
charmonium, which had a long story of a debate whether it will survive
in QGP or not. The answer depends on which
is the effective potential: in contrast to many earlier works
we pointed out~\cite{SZ_rethinking} that one has
to remove the entropy term and
use the ``energy'' potentials $V(T,r)=F-TdF/dT$, not the free energy
ones $F(T,r)$ which are directly measured from Wilson/Polyakov lines. 
This leads effectively to deeper
potentials
and better binding, so $J/\psi$ survives till $T=(2-3)T_c$.
This was confirmed by direct calculation of spectral densities
by maximal entropy method \cite{charmonium}. It also nicely correlates
well with surprisingly
small $J/\psi$ suppression observed at RHIC, where $T<2T_c$.

 It was then pointed out in \cite{SZ_bound} that also
multiple binary {\em colored} bound states should exist in the
same $T$ domain. Since QGP is a deconfined phase, there is nothing
wrong with that, and the forces between say  singlet $\bar q q$
and octet $q g$ quasiparticle pairs are about the same. 
Liao and myself~\cite{LS} have also found survival of the 
s-wave baryons ($N,\Delta...$) at $T<1.6T_c$. 

 A particularly interesting objects are
 {\em multibody}~\cite{LS} bound states, such as
``polymer chains''. Those can be ``open strings'' $\bar
q g g ..g q$ or closed chains of gluons (e.g. very robust $ggg$
state we studied). Their binding {\em per bond} was proven
 to be the same as
for light-quark mesons, and both are
bound till about 1.5$T_c$ or so. They have interesting
AdS/CFT analogs (see below) and they also can be viewed as 
precursor to the formation of the QCD strings from the 
deconfined phases.

 A curve of marginal stability (CMS) is not a thermodynamic
 singularity
but it often indicates a change of physics. Zahed and
 myself~\cite{SZ_rethinking} argued that 
 resonances can strongly enhance transport
 cross section near multiple CMS's and thus explain small viscosity.
Rapp and van Hees~\cite{Rapp_vanHees} studied $\bar q c$
 resonances, and found  enhancement of charm
 stopping.

 Similar phenomenon 
does happen for ultracold trapped atoms, which are extremely dilute
but due to
Feshbach-type resonances at which
the scattering length $a\rightarrow \infty$ they behave like very
good liquids with small viscosity, see\cite{Gelman:2004fj}. 

One notable colored bound state is a diquark $qq$, the main player 
in color superconductivity at high density and low $T$. 
Diquarks are weaker bound
than mesons\footnote{Due to extra 1/2 in color Casimirs.} and are
expected to melts right above the deconfinement transition.
In my recent paper\cite{Shuryak:2006ap}
I argued that may bring color superconductivity into strongly coupled
regime as well. The usual BCS theory of superconductors
are then inapplicable: it is also weak coupling theory.

 Fortunately
superfluidity of
 ultracold fermionic strongly coupled
 atoms have been studied recently experimentally.
The system is known to enter the  universal
strongly coupled regime as their scattering length $a$ gets
 large, and therefore it is possible to use 
some universal properties to get such properties as
the slope of the critical line of color superconductivity,
 $dT_c/d\mu$. I also deduced limitations on 
the critical temperature of color superconductivity itself
and conclude that it is limited by $T_{CS}<70\, MeV$.

\section{AdS/CFT correspondence at finite $T$}

{\bf Thermodynamics} of the CFT plasma was studied started from the
early work\cite{thermo}, its  result 
is that the free energy (pressure) of a plasma is
\be F(g,N_c,T)/ F(g=0,N_c,T)=[(3/4)+O((g^2N_c)^{-3/2})] \ee 
which  compares well with the
 lattice value\footnote{ Not too close to $T_c$, of course, but
 in the
``conformal domain'' of $T= few\,T_c$, in which
$p/T^4$ and $\epsilon/T^4$ are constant.} of about $0.8$.

{\bf Heavy-quark potentials} in vacuum  and then
at finite $T$ \cite{AdS_pot}
 were calculated by calculating the configuration
of the static string, deformed by gravity into the 5-th dimension.
Let me write the result schematically as
\be V(T,r,g)\sim  {\sqrt{g^2 N_c}\over r} exp(-\pi T r)  \ee 
The
 Debye radius at  strong coupling is unusual: unlike in pQCD it has no
 coupling constant.
Although potential depends on distance $r$ still
as in the Coulomb law, $1/r$ (at $T=0$ it is due to conformity),
 it is has a notorious square root of the coupling. 
Semenoff and Zarembo \cite{Semenoff:2002kk} noticed that summing ladder diagrams
one can explain $\sqrt{g^2 N_c}$, although
not a numerical constant. Zahed and myself~\cite{SZ_CFT}
pointed out that  both static charges are color correlated 
during a parametrically small time  $\delta t\sim r/{(g^2
  N_c)^{1/4}}$: this explains~\cite{Klebanov:2006jj} why a field of the dipole   
is $1/r^7$ at large distance\cite{Callan:1999ki}, not  $1/r^6$. 
Debye screening range can also be explained by resummation of thermal
polarizations~\cite{SZ_CFT}.

 Zahed and myself~\cite{SZ_spin} had also discussed
the velocity-dependent
forces , as well as spin-spin and spin-orbit ones, at strong coupling.
Using
ladder resummation for non-parallel
Wilson lines with spin they  concluded that all of them
join into one common square root
\be V(T,r,g)\sim \sqrt{(g^2 N_c)[1-\vec v_1 *\vec v_2+(spin-spin)+(spin-orbit)]}/ r   \ee 
Here $\vec v_1,\vec v_2$ are velocities of the quarks: 
and the corresponding term is a strong coupling version of Ampere's
interaction between two currents\footnote{Note that in a 
quarkonium  their scalar product is negative,
increasing attraction.}. No results on that are known from a gravity
side, to my knowledge.

{\bf Bound states} Zahed and myself~\cite{SZ_CFT} looked for
 heavy quarks   bound states, using a  Coulombic potential with 
Maldacena's $\sqrt{g^2 N_c}$ and
 Klein-Gordon/Dirac eqns. There is no problem with states
 at large orbital momentum $J>>\sqrt{g^2N_c}$, otherwise one has 
the famous  ``falling on a center'' solutions\footnote{Note that all
 relativistic corrections mentioned above  cannot prevent it
from happening.}: we argued that
a significant density of bound states develops, at all energies, from zero to
$2M_{HQ}$.  

And yet, a study of the gravity side~\cite{Kruczenski:2003be} 
found that there is no falling. In more detail, the
Coulombic states at large $J$ are supplemented by two more
families: Regge ones with the mass $\sim M_{HQ}/(g^2N_c)^{1/4}$ and
 the lowest $s$-wave states
(one may call $\eta_c,J/\psi$)  with even smaller
masses  $\sim M_{HQ}/\sqrt{g^2  N_c}$.
 The issue of ``falling'' was further discussed
 by Klebanov, Maldacena and Thorn~\cite{Klebanov:2006jj} for
 a pair of static quarks: they calculated the
 spectral density of states
  via a semiclassical quantization of  string vibrations.
They argued  that their corresponding density of states
should appear at exactly the same
critical
coupling as the famous ``falling'' in the Klein-Gordon eqn..

AdS/CFT also has  multi-body states  similar
to   ``polymeric
chains'' $\bar q.g.g... q$ discussed above. For
the endpoints being static quarks and the intermediate
gluons  conveniently replaced by adjoint scalars,
 Hong, Yoon and Strassler~\cite{Hong:2004gz} have studied
such states and even their formfactors.

{\bf Transport properties} of the CFT plasma
was a subject of recent breakthroughs 
\footnote{The works which appeared
between the conference and the time when this summary is written
are included.}. The (already famous) work
  by Polykastro, Son and Starinets\cite{PSS}
 have calculated  viscosity
(at infinite coupling)
 $ \eta/s=>1/4\pi$
which is in the ballpark of
 the empirical RHIC value. 
%They obtained this result from Kubo formula and 
%also from observing a sound pole in stress tensor correlator.
It taught us that
% important conclusion: 
gravitons in the bulk 
at large distances are dual
to phonons on the brane. Dual to
a viscous sound absorption is thus interception
of gravitons by the black hole.

Heavy quark diffusion constant
has been calculated by Casalderrey-Solana and
Teaney~\cite{Casalderrey-Solana:2006rq}: their result is
\be D_{HQ}={2 \over \pi T \sqrt{g^2 N_c}}  \ee
which is parametrically smaller than 
an expression for the momentum diffusion
$D_p=\eta/(\epsilon+p)\sim 1/4\pi T$.
This work is methodically quite different from others in that Kruskal
coordinates are used, which allows to consider the inside
of the black hole and $two$ Universes (with opposite time directions)
simultaneously, see Fig.3a. This is indeed necessary
\footnote{One such problem is evaluation of the so called $\hat q$
parameter: two lines of the loop  should 
also belong to $two$ different Universes, not one as assumed in
\cite{Liu:2006ug}. It remains unknown  whether
similar calculation in Kruskal geometry
would produe the same result or not. }
 in any problems when
a $probability$ is evaluated, because that contains
both an amplitude and
a conjugated amplitude at the same time. 
  
Jet quenching studies\cite{Sin:2004yx,Herzog:2006gh,Gubser:2006bz,Buchel:2006bv,Sin:2006yz} have been reported by L.Yaffe, see his talk for
details.
The result for the drag force is
\be {dP\over dt}= -{\pi T^2\sqrt{g^2 N_c} v \over  2\sqrt{1-v^2}} \ee
Quite remarkably, the Einstein relation which 
relates  the heavy quark diffusion constant (given above)
to the drag force
is actually fulfilled, in spite of quite different gravity settings
shown in Fig.3, a and b.

This result is valid only for quarks heavy enough $M>M_{eff}\sim
\sqrt{g^2 N_c} T$ and is
obtained in a stationary setting, in which
a quark is dragged with constant by ``an invisible hand'' via some rope
through QGP, resulting in constant production of a string length per
time, see Fig.3b . I have borrowed it from the paper by Friess et al
\cite{Friess:2006aw}, who have made the next
(and technically much more difficult) step, namely solving
the Einstein equation with this falling string as a source
and found corrections to the metric
$h_{\mu\nu}$ and thus the matter stress tensor
on the brane. Quite remarkably, when they analyzed harmonics
of this stress at small momenta they have seen the ``conical flow''!
And, as one can see from plots for ``subsonic'' $v<1/\sqrt{3}$,
the Mach cone disappear in this case, as argued in \cite{Antinori:2005tu}.

\begin{figure}
\includegraphics[width=4.5cm]{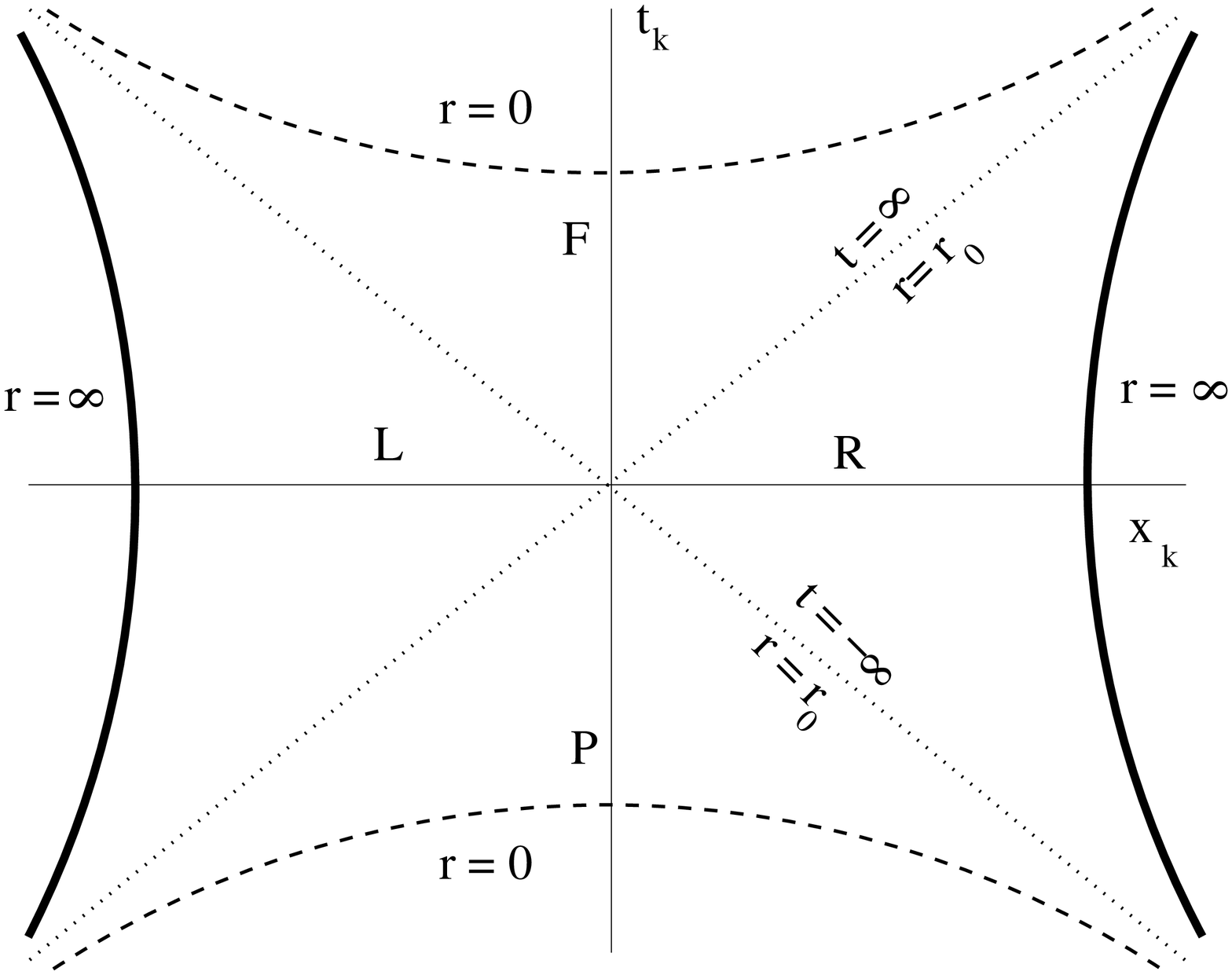}
\includegraphics[width=5.5cm]{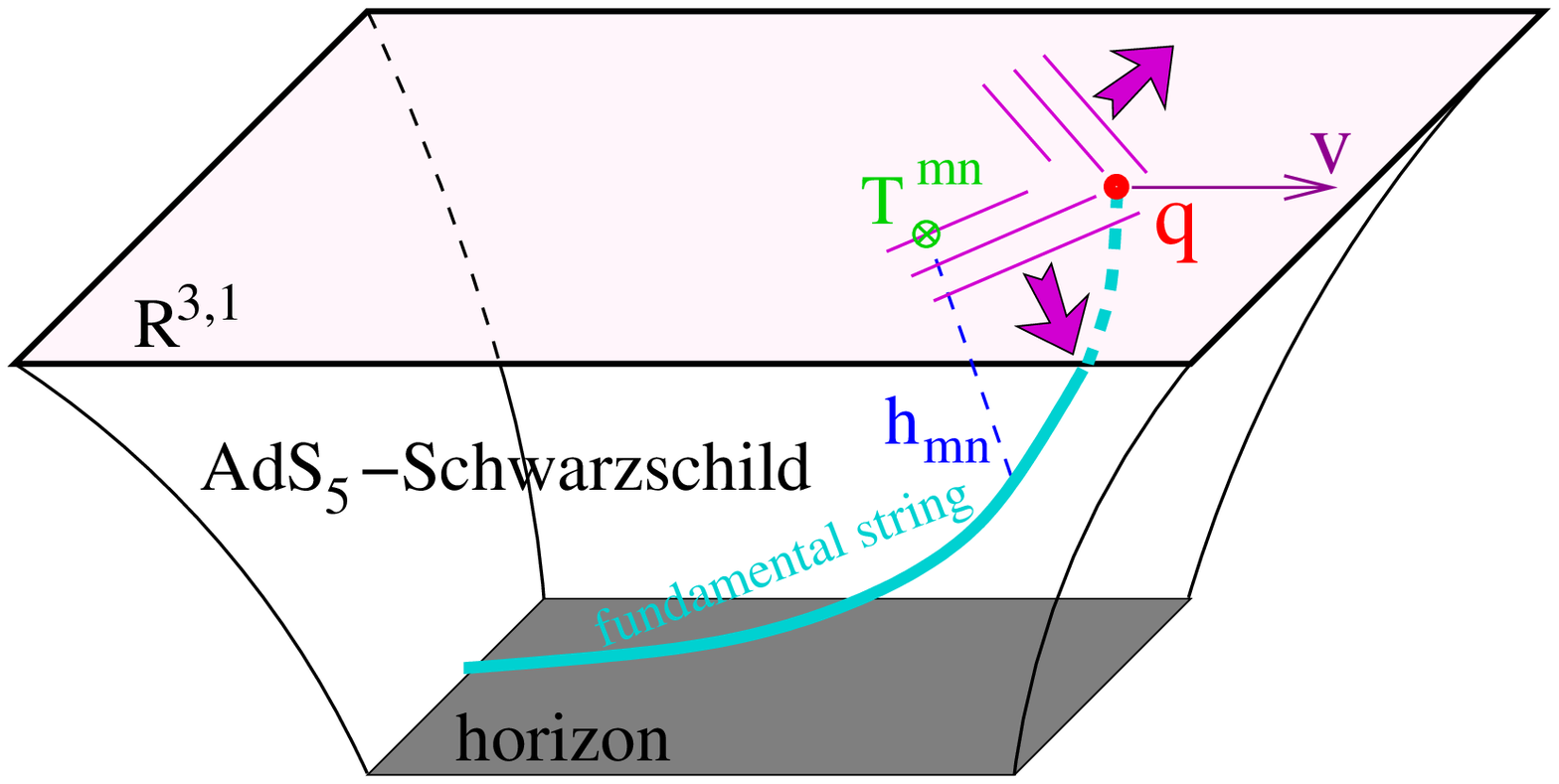}
 \caption{\label{fig_wake}
(a) (from \protect\cite{Casalderrey-Solana:2006rq}): In Kruskal coordinates one can study two Universes
at the same time, shown right and left, and the evaluated
Wilson line contains static quarks on their boundaries.
(b) (from \protect\cite{Friess:2006aw})
The dragged quark trails a string into the five-dimensional AdS
 bulk, representing color fields sourced by the quark's fundamental
 charge and interacting with the thermal medium. The back gravity
 reaction describes how matter flows on the brane.
 }
 \end{figure}

% \begin{figure}
%  \centerline{\includegraphics[width=5in,angle=270]{BoomFig.eps}}
%  \caption{Contour plots of $K_\perp |Q^K_E|$ for various values of $v$ at low momenta.  The green line shows the Mach angle.  The red curve shows where $K_\perp |Q^K_E|$ is maximized for fixed $K = \sqrt{K_1^2+K_\perp^2}$.  The blue curves show where $K_\perp |Q^K_E|$ takes on half its maximum value for fixed $K$.}\label{fig:BoomFig}
% \end{figure}

  The ultimate goal would be a complete
 ``gravity dual'' to the
whole
RHIC collision process, in which thermalization and subsequent
hydro explosion will be
described via dynamical  production of a black hole,
as emphasized by Nastase~\cite{Nastase}. 
 Sin, Zahed and myself~\cite{Shuryak:2005ia}
further argued that exploding and cooling
fireball on the brane is dual to
 departing  black hole,  formed by collision debris 
falling into the AdS center.
Janik and Peschanski~\cite{JP},
 found that  the Bjorken expansion 
can be mapped into a metric with a departing $horizon$\footnote{However
they solve vacuum Einstein eqns without any matter: their departing
horison is due to acausal time dependence of the 
central black hole (the stack of branes).}.

\section{AdS/QCD}
 ``Holographic'' ideas have been also used
 for theories more resembling QCD.
 There are famous papers on it; my favorite is Sakai-Sugimoto
model\cite{Sakai:2005yt} with light quarks,
deconfinement and chiral symmetry
restoration. The closest to RHIC physics is nice
 paper by Peeters et al\cite{Peeters:2006iu} which document
how light mesons (e.g. $\rho$)
  get $T$-dependent masses after they  survive deconfinement.

Quite intriquing is also a
``bottom-up'' approach, in which the
violations of comformity of AdS/CFT is introduced explicitly.
 Confinement forbids 
QCD phenomena from exploring large distances. In the 
now popular ``holographic'' language, it means that all object
we study are somehow prevented
from going too far into the 5-th coordinate $z$ into the IR.
First attempts to  model confinement (used mostly for QCD spectroscopy)
 used just such a  cut-off,  at some $z=z_0$. 

 Recently three groups  suggested different arguments that
  confinement induces a $quadratic$ potential in $z$.
Karch et al~\cite{KKSS} have argued that this is needed to get
correct dependence of the Regge trajectories on particle spin $S$ and
principal quantum number $n$.
In my paper~\cite{Shuryak:2006yx} the probe for confinement are
  instantons, and it was first argued that their size should be
identified to the 5-th coordinate $z$.
The proposed potential consists of two parts,
related to asymptotic freedom\footnote{I found it remarcable that
both other groups igore asymptotic freedom and deviations from
conformity in UV. The corresponding coupling constants
for any operator is related to its perturbative anomalous dimension.
} and confinement
\be V_{eff}(z)=V_{AF}+V_{conf}=-\beta_0 log(z)+ 2\pi\sigma z^2\ee
where $\beta_0=(11/3)N-(2/3)N_f$ and 
the coefficient of the quadratic term was proposed previously
 in \cite{Shuryak:1999fe}, it contains the string tension $\sigma$.
 The reason (and coefficient\footnote{It is similar but not the same
as the one proposed in \cite{KKSS}, which 
 should not surprise us, as the constant depends on how
strongly the object studied is coupled to confinement-related
condensates. For example, for glueball
Regge trajectories 
(including the Pomeron and the 2++ glueball)  the Regge slope 
is already significantly
different.
}
)
 for quadratic behavior is not ad hoc, but because a it is related with
 a VEV of a {\em dual superconductor}; it is in excellent agreement
with lattice datat on instanton size distribution.

Andreev and Zakharov~\cite{Andreev_Zakharov} put the quadratic
potential into metric, and calculated a number of string-based
potentials and spatial Wilson lines. However if it is in metric,
wrong Regge trajectories follows\cite{KKSS} and also one cannot
have non-universal coupling just mentioned: so I think it should
be put into some extra potential instead.  

\section{(Post)Confinement and monopoles}
Here comes an old question: is there any progress in understanding
confinement, as well as the deconfinement transition region, at
$T\approx T_c$?

Recent lattice data have revealed a puzzling behavior
of static $\bar Q Q$ potentials, which I call
``postconfinement''. At $T=0$ we all know that a potential between
heavy quarks is a sum of the Coulomb and a confining $\sigma(T=0) r$ 
potential. At deconfinement $T=T_c$ the Wilson or Polyakov lines with
a static quark pair
has  vanishing string tension; but this is the free energy
$exp(-F(T,r))=<W>$. Quite shockingly, if one calculates the $energy$
or $entropy$ separately (by $F=E-TS$, $S=-\partial F/\partial T$)
one finds \cite{potentials}
 a force between $\bar Q Q$ to be more than twice
  $\sigma(T=0)$ till rather large distances. The total energy 
added to a pair is surprisingly large
reaches about $E(T=T_c,r\rightarrow\infty)=3-4\, GeV$, and the
entropy as large as $S(T=T_c,r\rightarrow\infty)\sim 10$. Since this energy of ``associated matter''
is about 20 times larger than $T$, any separation of two static quarks
must be extremely suppressed by the Boltzmann factor exp(-E/T).
(As $T$ grows, this phenomenon disappears,
and thus it is obviously related to the phase transition itself.)

Where all this energy and entropy may come from in the
 deconfined phase? 
It can only be long and complicated QCD string 
connecting two static quarks. 
We already mentioned that such strings can be explained by 
a ``polymerization'' of gluonic quasiparticles in sQGP.

Let us now add a twist to this story related with magnetic
excitations,
the monopoles\footnote{Recall that they appear naturally if
there is an explicit Higgs VEV breaking of the color group.
We cannot discuss in detail a QCD setting: the reader may
simply imagine a generic finite-$T$ configuration with
a nonzero mean $<A_0>$, an  adjoint Higgsing leaving
$N_c-1$ U(1) massless gauge fields. These U(1)'s corresponds to
magnetic charges of the monopoles. In AdS/CFT language one may simply
considered $N_c$ branes to be placed not at exactly the same point
in the orthogonal space.}.
 According to t'Hooft-Mandelstamm scenario, confinement is supposed to be due to monopole
condensation. Seiberg-Witten solution for the \cal{N}=2 SYM is an
example of how it is all supposed to work: it has taught us that
as one approaches the deconfinement transition
the electrically charged particles -- quarks and gluons --
are getting heavier while monopoles gets lighter and more numerous.
Although I cannot go into details here, we do have hints from
lattice studies of monopoles and related observables that
this is happening in QCD as well.
 
Let us now think what
all of it means for the sQGP close to $T_c$. Even at classical level, it means that
one has a plasma with both type of charges -- {\it electric and magnetic} --
at the same time, with the former dominant at large $T$ and
the latter dominant close to $T_c$.  

A binary dyon-dyon systems have been studied before , but not
manybody ones.
The first numerical studies of such systems
(by molecular dynamics) are now performed by
(Stony Brook student) Liao and myself \cite{LS_monopoles}. We found
that a monopole can be trapped by an electric static dipole, both
classically and quantum mechanically. We also found that classical
gas of monopoles leads to electric 
flux tubes\footnote{Those are dual to magnetic flux tubes in
solar classical plasmas.} because monopoles scatter from the electric
flux tube back into plasma, compressing it.
Whether monopoles  are condensed or not is not crucial.

Are ther bound states of electric and magnetic quasiparticles?
Yes, there are a lot of them. A surprize is that
even  finite-$T$ instantons
can be viewed as being made of $N_c$ selfdual dyons~\cite{Kraan:1998kp},  attracted to
each other  pairvise, electrically and magnetically.
  Not only such baryons-made-of-dyons have the same moduli space
as instantons,  the solutions can be obtained
vis very interesting AdS/CFT brane
construction \cite{Lee:1997vp}. Many more exotic bound states
of those are surely waiting  to
be discoverd. 

%%%%%%%%%%%%%%%%%%%%%%%%%%%%%%%

\end{document}